\documentclass[aps,floatfix,showpacs,twocolumn,pra]{revtex4}
\usepackage[dvips]{graphicx}
\usepackage{amsfonts}

\newcommand{\bra}[1]{{\langle #1 |}}
\newcommand{\ket}[1]{{| #1 \rangle}}
\newcommand{\braket}[2]{\langle #1 | #2 \rangle}
\newcommand{\tr}{\mathrm{tr}}

\begin{document}

\title{Entanglement Across a Transition to Quantum Chaos}
\author{Carlos Mej\'{\i}a-Monasterio$^{(a)}$} 
\email{carlos.mejia@uninsubria.it}
\homepage{http://scienze-como.uninsubria.it/complexcomo}
\author{Guliano Benenti$^{(a,b,c)}$}
\email{giuliano.benenti@uninsubria.it}
\author{Gabriel G. Carlo$^{(a,b)}$}
\email{gabriel.carlo@uninsubria.it}
\author{Giulio Casati$^{(a,b,c)}$} 
\email{giulio.casati@uninsubria.it} 
\affiliation{$^{(a)}$Center for Nonlinear and Complex Systems,  
Universit\`a degli Studi dell'Insubria, via Vallegio 11, Como 22100, Italy}
\affiliation{$^{(b)}$Istituto Nazionale per la Fisica della Materia,
Unit\`a di Como, via Vallegio 11, Como 22100, Italy}
\affiliation{$^{(c)}$Istituto Nazionale di Fisica Nucleare,
Sezione di Milano, Via Celoria 16, 20133 Milano, Italy}

\date{\today}

\begin{abstract}
We study the  relation between entanglement and quantum  chaos in one-
and two-dimensional spin-$1/2$ lattice models, which exhibit mixing of
the noninteracting eigenfunctions and transition from integrability to
quantum chaos. Contrary to what  occurs in a quantum phase transition,
the onset of  quantum chaos is not a property of  the ground state but
take  place  for  any  typical  many-spin  quantum  state.   We  study
bipartite and  pairwise entanglement measures, namely  the reduced Von
Neumann entropy and the  concurrence, and discuss quantum entanglement
sharing.  Our results suggest that the behavior of the entanglement is
related  to  the mixing  of  the  eigenfunctions  rather than  to  the
transition to chaos.
\end{abstract}


\maketitle

\section{Introduction}
Quantum  entanglement  has been  identified  as  a  key ingredient  in
quantum  communication  and information  processing.   The content  of
entanglement in  a particular  system is considered  as a  resource to
perform several tasks in a more efficient and more secure way than any
other  classical   method  \cite{chuang00,benenti04}.   For  instance,
quantum teleportation protocols require to share a maximally entangled
state between  the sender  and the receiver.  In the field  of quantum
computation, it has been found that for the case of quantum algorithms
operating on  pure states,  the presence of  multipartite entanglement
between the components constituting a quantum processor is a necessary
condition to achieve an exponential speedup over classical computation
\cite{linden02}.

On the  other hand, for the  operability and stability  of any quantum
computer, the entanglement  can also play the role  of the property to
be  minimized.   The  unavoidable  entanglement  between  the  quantum
processor and the environment is  one of the most important sources of
noise and, therefore, of  computational errors.  The understanding and
control of noise  in quantum protocols is clearly  needed to implement
any              reliable              quantum             computation
\cite{paz,zurek,song,carlo,rossini03,facchi}.   Even  when  a  quantum
processor is  ideally  isolated from  the  environment, {\it i.e.}, in
situations where  the decoherence time of the  processor is very large
as compared to  the computational time  scales, the operability of the
quantum computer is   not yet guaranteed \cite{georgeot00}.    Indeed,
also the presence of device  imperfections hinders the  implementation
of any quantum computation.  A quantum computer is a quantum many-body
system.  The interaction   between  the qubits composing  the  quantum
registers of the computer is needed to produce the necessary amount of
entanglement.  Moreover, device imperfections like small inaccuracy in
the coupling  constants  induce errors.  Above  a certain imperfection
strength  threshold  ({\it chaos    border}),  quantum chaos   sets in
\cite{georgeot00,flambaum,benenti01,berman01}.  In  such    a  regime,
exponentially many states of the computational basis are mixed after a
chaotic time scale.  This sets an upper time limit to the stability of
a generic superposition  of states coded in  the quantum computer wave
function.  A  necessary requirement  for quantum  computer operability
and fault  tolerant computation schemes  is the possibility to operate
many quantum gates inside the chaotic time scale.

For the  case of many-body systems,  the transition to  chaos has been
studied  for  fermions and  bosons  (see  e.g. \cite{many-bodies}  and
references  therein),  and   particularly  for  lattice  spin  systems
\cite{spin-systems}.  In fermionic  systems with two-body interactions
it  has been  found that,  if  the interaction  strength exceeds  some
critical value, fast  transition to chaos occurs in  the Hilbert space
of many-particle  states \cite{izrailev99}.  This  is commonly studied
in terms  of the transition between the  different spectral statistics
of integrable  and chaotic systems.   For systems with a  finite size,
this  transition is  smooth  and  only a  crossover  border where  the
transition occurs can be identified.  The question whether this smooth
transition  becomes sharp in  the thermodynamic  limit is  still under
debate.  However, in some cases  a sharp transition to chaos is found,
{\it    e.g.},    in     the    three-dimensional    Anderson    model
\cite{3D-anderson}.

In a different context, the  behavior of quantum entanglement across a
quantum  phase  transition  has  recently  attracted  much  attention.
Quantum phase transitions (QPT) consist in a qualitative change in the
ground  state  of the  system  as  some  control parameter  is  varied
\cite{sachdev}.  Unlike classical phase transitions, QPT occur at zero
temperature and  the fluctuations developed at the  critical point are
fully quantum.  These fluctuations dominate the behavior of the system
near  the  critical  point,   where  correlations  are  long-range  in
character.  It has been recently  pointed out that the genuine quantum
character  of  quantum  phase   transitions  is  due  to  entanglement
\cite{osborne01}.  It  has also been  argued that the ground  state of
the   system   is   strongly   entangled   at   the   critical   point
\cite{osborne01,volya03}.   Therefore,  the  behavior of  entanglement
across  QPT is  particularly interesting  for quantum  computation and
communication, where a maximization  of the content of entanglement is
desirable.  The study of the relation between QPT and entanglement has
been focused  on the possible  universal behavior of  the entanglement
content                at                the                transition
\cite{fazio02,osborne02,bose02,vidal03,gu03,glaser03,korepin04,mosseri04,cirac04}.
In this  context, it  has been shown  that in different  model systems
quantities     like    the     derivative    of     the    concurrence
\cite{fazio02,osborne02}       and      Von       Neumann      entropy
\cite{vidal03,latorre-S} present critical  behavior at the transition.
The  dependence of  entanglement on  disorder and  its  interplay with
chaos has also been studied \cite{simone,santos03,emary04}.  Moreover,
the evolution of entanglement in quantum algorithms simulating quantum
chaos          has         been          recently         investigated
\cite{bettelli03,caves,lahiri03,hu03-1}.

The  aim of  this paper  is to  characterize the  behavior  of quantum
entanglement  in \emph{non-integrable}  systems when  a  transition to
quantum  chaos occurs.   We are  interested in  understanding  how the
entanglement  content  behaves in  transitions  from integrability  to
chaos.  We  would like to  stress that, differently from  the previous
studies                             on                             QPT
\cite{fazio02,osborne02,bose02,vidal03,gu03,glaser03,korepin04,mosseri04,cirac04},
the transition to chaos is not only a property of the ground state but
takes place for any typical many-body state.  We numerically study two
lattice models  of interacting  many spins that  show a  transition to
chaos.  They  have previously  been studied as  models of  the quantum
computer  hardware  \cite{georgeot00,flambaum,benenti01,berman01}.  In
both models, the transition to chaos  is driven by the strength of the
interaction  between the  spins.  We  consider bipartite  and pairwise
entanglement  measures   and  focus   on  the  relation   between  the
entanglement  and the  onset  of chaos.   We  focus our  study on  the
eigenstates  corresponding to the  center of  the spectrum,  where the
many-body density of states is larger and therefore quantum chaos sets
in at small interaction  strengths. Nevertheless, also the behavior of
the  other  parts  of  the   spectrum  is  discussed.   We  use  exact
diagonalization  techniques  to  obtain  all the  eigenstates  of  the
considered  spin  models.  Therefore,   we  are  limited  to  consider
relatively small  system sizes, from  which the study of  any possible
finite-size scaling for the behavior  of the entanglement at the chaos
border is  out of reach.   Nevertheless, we discuss, at  a qualitative
level,  the  similarities  and  differences between  the  behavior  of
entanglement across a QPT and at  the onset of quantum chaos.  We show
that the dependence of pairwise entanglement on the size, distance and
range of the interactions can be understood in terms of the sharing of
entanglement among the different  parties of the system.  Furthermore,
we demonstrate  that the  behavior of entanglement  is related  to the
mixing of noninteracting eigenfunctions  rather than to the transition
to chaos.

This  paper is organized   as follows.  In Sec.~\ref{sec:measures}  we
review the  measures of entanglement that  we will  use.  The measures
that    signal the  onset   of   quantum    chaos  are reviewed     in
Sec.~\ref{sec:chaos}.  In Sec.~\ref{sec:examples}  we define the  spin
lattice  models investigated  in   this  paper and discuss   numerical
results on  their entanglement content  and their transition to chaos.
In Sec.~\ref{sec:conclusions} we present our final remarks.

\section{Measures of Entanglement}
\label{sec:measures}

A  pure state  $\ket{\psi}$ is  said to  be separable  if for  a given
partition of its Hilbert space  ${\mathcal H} = {\mathcal H}_A \otimes
{\mathcal H}_B$  it can  be written as  $\ket{\psi} =  \ket{a} \otimes
\ket{b}$.
Here  $\ket{a}$ and  $\ket{b}$  are vectors  residing  in the  Hilbert
subspaces  ${\mathcal H}_A$ and  ${\mathcal H}_B$,  respectively.  The
pure state $\ket{\psi}$ is entangled if it is not separable.

\subsection{Von Neumann entropy}
\label{sec:entropy}

Pure bipartite  entanglement is measured  in terms of the  reduced Von
Neumann entropy  $S$.  For a pure  state the reduction  of its density
matrix $\rho = \ket{\psi}\bra{\psi}$ is obtained through partial trace
of one  the two partitions  as $\rho_A = \tr_B\rho$  or, equivalently,
$\rho_B = \tr_A\rho$. Then, $S$ is defined as
\begin{equation} \label{eq:entropy}
S = S_A = S_B = -\tr_B (\rho_B \log \rho_B) \ .
\end{equation}
The   Von  Neumann   entropy  provides   an  unambiguous   measure  of
entanglement for  a bipartite system in  an overall pure  state. For a
separable state  $S=0$ while for  a maximally entangled  state $S=\log
{\mathcal  N}$,  where  ${\mathcal  N}=\min({{\mathcal  N}_A,{\mathcal
N}_B})$,   with  ${\mathcal   N}_A={\rm   dim}({\mathcal  H}_A)$   and
${\mathcal N}_B={\rm  dim}({\mathcal H}_B)$.  In what  follows we will
take  the logarithm in  Eq.~(\ref{eq:entropy}) base  $2$. Thus,  for a
many-qubit system, the maximum value  that the Von Neumann entropy can
take is equal to the number of qubits that have not been traced out to
obtain the reduced density matrix.

\subsection{Concurrence and entanglement of formation}
\label{sec:concurrence}

For the case  of mixed states, the Von Neumann entropy  is no longer a
good measure of entanglement. If  we consider a bipartite system on an
overall  mixed state, then  each subsystem  can have  non-zero entropy
even if there is not  any entanglement \cite{wootters01}.  In order to
measure the bipartite entanglement of  a mixed state we shall consider
the  so-called  entanglement   of  formation  $E_F$  \cite{bennett96}.
Starting  from  a mixed  state  with  density  matrix $\rho  =  \sum_i
p_i\ket{\psi_i}\bra{\psi_i}$,  $E_F(\rho)$ is  defined as  the average
entanglement of the pure states  of a given decomposition of the mixed
state $\rho$, minimized over all its possible decompositions:
\begin{equation} \label{eq:E_F}
E_F(\rho) = \min_{\{p_i,\psi_i\}} \sum_i p_i E(\ket{\psi_i}) \ .
\end{equation}
Here $E(\ket{\psi_i})$ is the amount of entanglement of the pure state
$|\psi_i\rangle$, measured,  as discussed in  the previous subsection,
by the reduced Von Neumann entropy.

Eq.~(\ref{eq:E_F})  is  operationally difficult  to  handle, since  it
involves an  extremal condition.  However,  for the case  of two-qubit
systems,  $E_F$ can  be expressed  in terms  of a  much  more amenable
quantity, the so-called concurrence $C$ \cite{wootters98}. We have
\begin{equation} \label{eq:E_F(C)}
E_F(\rho) = h\left(\frac{1}{2}\left[1 + \sqrt{1-C(\rho)^2} \right]\right) \ ,
\end{equation}
where $h(x)$ is the so-called binary entropy function defined by
\begin{equation} \label{eq:binary_h}
h(x) = -x\log_2x -(1-x)\log_2(1-x) \ .
\end {equation}
The concurrence $C(\rho)$ of the two-qubit state $\rho$ is defined as
\begin{equation} \label{eq:C}
C(\rho) = \max\{0,c_\lambda\} \ ,
\end{equation}
where $c_\lambda = \lambda_1 - \lambda_2 - \lambda_3 - \lambda_4$, the
$\{\lambda_i\}$  being the  square  roots of  the  eigenvalues of  the
matrix $\rho(\sigma_y\otimes\sigma_y)\rho^*(\sigma_y\otimes\sigma_y)$,
in decreasing order, and $\sigma_y$  a Pauli matrix. Note that in this
definition the complex conjugation is taken in the computational basis
$\{|00\rangle,|01\rangle,|10\rangle,|11\rangle\}$,  where  $|0\rangle$
and   $|1\rangle$   are   the   eigenstates   of   $\sigma_z$.    From
Eq.~(\ref{eq:E_F(C)}) we  see that  $E_F$ depends monotonously  on the
concurrence, which  takes values between $0$ for  separable states and
$1$ for maximally  entangled states. Moreover, it is  easy to see that
$c_\lambda$  take  values  in  $[-1/2,1]$.   Thus, it  is  clear  from
Eq.~(\ref{eq:C}) that  a state is  separable if $c_\lambda \le  0$ and
entangled otherwise.

Other measures for pairwise entanglement exist.  Among them we mention
the entanglement  of  distillation \cite{distillation}, the negativity
\cite{negativity}    and              the          relative    entropy
\cite{relative_entropy}. All these  quantities are related in one  way
or  another to the concurrence. Therefore,  we  have chosen to present
our results  in terms  of  the concurrence.   Nevertheless, we mention
that  we  have  also   measured the   negativity and  found   that  it
essentially gives the same results as the concurrence.

Finally, for the case of multipartite entanglement the problem is much
more  subtle.   Different measures  of  multipartite entanglement have
been  proposed,   giving    different  results,    even  qualitatively
\cite{multipartite}.
Given the  state  of  affairs,  we  will  limit
ourselves to  discuss  the existence of  multipartite  entanglement in
terms   of   the qualitative information that     can  be extracted by
comparing the averaged Von Neumann entropy for subsystems of different
sizes.

\section{Transition to Quantum Chaos}
\label{sec:chaos}

Random  Matrix Theory (RMT)  was  introduced to  describe the spectral
properties of complex heavy nuclei.   The key  idea  behind RMT is  to
replace the full physical description of the Hamiltonian by a suitable
statistical representative of  its   symmetry group \cite{rmt}.    The
statistical spectral fluctuations  of  almost any  complex Hamiltonian
were    found  to be  described  by  a  few  classes  of random matrix
ensembles.  This approach  turned out to be  very successful.  The RMT
analysis has been applied  to many fields of  physics such as  nuclei,
atoms,   molecules,  quantum  dots,  quantum  billiards and  many-body
systems among others
\cite{brody81,guhr98,varena99,haake,spin-systems,many-bodies}.  In the
early 1980's   it was   conjectured  that  the  quantum   versions  of
integrable and chaotic classical  systems were described by  different
classes of  random ensembles  \cite{bohigas84,casati80}.   Since then,
RMT has been successfully applied to describe the emergence of quantum
signatures of chaos.

The global  manifestation of  the onset  of chaos in   quantum systems
consists of a very complex structure of the quantum  states as well as
in spectral  fluctuations that  are  statistically described   by  RMT
\cite{rmt}.  Let us   focus  on many-particle systems  with   two-body
interaction as this is  the nature of the  model systems that we study
in this paper.  For this kind of systems it has been found that, under
very general conditions,  if  the  interaction strength  exceeds  some
critical value, fast transition  to chaos occurs  in the Hilbert space
of many-particle states.

To  be  more precise  let  us  consider  a generic  perturbed  quantum
many-body system.  The Hamiltonian can be split into two parts:
\begin{equation} \label{eq:H-m-b}
H = H_0 + V \ ,
\end{equation}
where $H_0$  corresponds  to the unperturbed original  Hamiltonian and
the  perturbation   $V$ to  an    interacting  term.  The  unperturbed
Hamiltonian $H_0$ is assumed integrable.   In other words, when  $V=0$
the existence of as many integrals of motion  as degrees of freedom is
assumed.     We will    take  the  unperturbed    eigenstates  of  the
non-interacting  Hamiltonian $H_0$  to  span the many-particle Hilbert
space.   In  this  basis,  when  the   interaction is  turned on,  the
eigenstates start  to mix.   The mixing can  be  described,  for small
interaction  strengths, by perturbation theory.  However, perturbation
theory  breaks  down  and  quantum chaos  sets   in when  the  typical
interaction matrix element between  directly coupled states becomes of
the order of their energy separation
\cite{izrailev99} (we  say that two  many-body states $|\psi_1\rangle$
and $|\psi_2\rangle$ are  directly connected if $\langle \psi_1  | V |
\psi_2  \rangle \ne  0$).  The  transition  to chaos  reflects in  the
statistical properties of the spectrum.

\subsection{Nearest neighbor level spacing distribution}
\label{sec:pds}

The  nearest neighbor   level  spacing  distribution   $P(s)$  is  the
probability density to find two adjacent levels at a distance $s$.

For an  integrable  system the distribution  $P(s)$ has  typically a
Poisson distribution
\begin{equation} \label{eq:pds_P}
P_{\rm P}(s)=\exp\left(-s\right) \ .
\end{equation}

In  contrast, in the  quantum chaos  regime, for  Hamiltonians obeying
time-reversal  invariance, the  nearest neighbor  spacing distribution
corresponds  to the  Gaussian Orthogonal  Ensemble of  random matrices
(GOE).  This distribution is  well-approximated by the Wigner surmise,
which reads
\begin{equation} \label{eq:pds_WD}
P_{\rm WD}(s) = \frac{\pi s}{2}\exp\left(-\frac{\pi s^2}{4}\right) \ .
\end{equation}
Note   that  the GOE   distribution  exhibits  the  so-called  ``level
repulsion'', {\it  i.e.}, the probability to  find close energy levels
is very small. This is in contrast to what  is observed for integrable
systems  which  exhibit level  clustering.   An example of Poisson and
Wigner distributions is provided in Fig.~\ref{fig:1D-pds}.

In Eqs.~(\ref{eq:pds_P})  and (\ref{eq:pds_WD}) the  level spacing $s$
is given in units of the mean level  spacing $\Delta$ that we have set
to $1$.

The transition to  quantum chaos may be  detected by the change of the
nearest neighbor spacing distribution $P(s)$ from Poisson  to GOE.  In
order to obtain a more  quantitative indication of this transition, it
is useful to compute the parameter
\begin{equation} \label{eq:gamma}
\gamma = \frac{\int_0^{s_0} \left[P(s) - P_{\rm WD}(s)\right] {\mathrm d}s}
{\int_0^{s_0} \left[P_{\rm P}(s) - P_{\rm WD}(s)\right] {\mathrm d}s} \ ,
\end{equation}
where $s_0  \approx 0.4729$  corresponds  to the lowest   $s$-value at
which       the         Poisson    [Eq.~(\ref{eq:pds_P})]  and  Wigner
[Eq.~(\ref{eq:pds_WD})] curves cross.  This parameter takes values $1$
and $0$ for the Poisson and Wigner distributions, respectively.

The distribution $P(s)$ describes the  behavior of the fluctuations at
energy scales  of the order of  $\Delta$. Therefore, $P(s)$ is a short
range correlation. The effects of the onset  of quantum chaos are also
seen in higher moments of the distribution of energy levels.  However,
we will limit  ourselves  to the  analysis   of $P(s)$ as a   spectral
signature of the transition to quantum chaos.

\subsection{Participation number}
\label{sec:PN}

The  effects  of the  onset   of chaos can   also be  observed in  the
eigenfunctions. The transition reflects in the degree of mixing of the
eigenfunctions  of  the   system.   However,  the  eigenfunction-based
measures  are   more    subtle.   This  is   because   the  mixing  of
eigenfunctions  is  a    basis-dependent  quantity.  Clearly,  if  the
eigenfunctions are expanded in their own basis,  they are not mixed at
all, independently of the fact that the distribution $P(s)$ is Poisson
or  GOE.  Nevertheless, for  Hamiltonians of the type (\ref{eq:H-m-b})
the  increase of typical eigenfunctions'   mixing with perturbation is
naturally obtained in  the   basis $\{\ket{i}\}$ of  the   unperturbed
Hamiltonian $H_0$ \cite{benet93}.

In the  basis    $\{\ket{i}\}$ the  mixing  (or,  equivalently,    the
\emph{delocalization})    of  a  given     eigenstate $\ket{\psi}$  is
customarily measured in     terms   of  the number      of  components
$\braket{i}{\psi}$ which   are  significantly different  from  zero: A
useful quantity to  measure the  degree of  delocalization of a  given
eigenstate is the so-called Participation Number (PN), defined
as
\begin{equation}\label{eq:PN}
\xi=\frac{1}{\sum_i|\braket{i}{\psi}|^4} \ .
\end{equation}
If the  state $\ket{\psi}$ is  maximally localized, all its components
are  zero    but  one, with  value     $|\braket{i}{\psi}|^2=1$ due to
normalization.   Therefore,  in the  localized  regime $\xi\approx 1$,
while $\xi$ increases with   increasing mixing.  In  the thermodynamic
limit, $\xi$ is unbounded. However, in the case of finite size systems
$\xi\le N$, where $N$ is the dimension of the  Hilbert space.  For the
GOE statistics, the  PN is upper bounded  by the value $\xi=N/3$, due
to  the  statistical  properties of  the chaotic states.

Even though, in general,  the transition from localized to delocalized
eigenfunctions occurs in  parallel with the transition from Poissonian
to Wigner spectra, this  is not always the  case.  As we will study in
the following sections, for  particular model systems the  spectra can
be  uncorrelated even  in situations  in  which the eigenfunctions are
delocalized.  It    is common to  term   this  situation as \emph{weak
chaos}.

\subsection{Chaos and entanglement}
\label{sec:borders}

It  is  worthwhile  to  discuss  what are  the  expectations  for  the
entanglement  content  in  the  nearly  integrable  and  fully  chaotic
situations.  Let us consider a many-particle system with a Hamiltonian
as in Eq.~(\ref{eq:H-m-b}) and  let $\mathcal{N}$ denote the dimension
of  its Hilbert  space.  In  a given  basis $\{\ket{n}\}$  the density
matrix  for an  eigenstate of  the  system, $\ket{\psi}  = \sum_n  c_n
\ket{n}$, writes as follows:
\begin{equation} \label{eq:dens-mat}
\rho_{nm} = \langle n \ket{\psi}\bra{\psi}m\rangle = c_n c_m^*  \ ,
\end{equation}
where $c_n = \braket{n}{\psi}$ are the components of the eigenstate in
the $\{|n\rangle\}$  basis.  As discussed in  the previous subsection,
the nearly integrable  case corresponds to  
a situation of  weak interaction.
This implies  that in the  basis of the unperturbed  Hamiltonian $H_0$
the   eigenfunctions   are  localized,   {\it   i.e.},  $c_n   \approx
\delta_{n,n^\star}$ for  some $n^\star\in[1,\mathcal{N}]$.  Therefore,
the density matrix will have  all entries nearly equal to zero, except
for   the   diagonal   matrix   element  $\rho_{n^\star   n^\star}   =
|c_{n^\star}|^2  \approx   1$.   At   the  other  extreme   of  strong
interaction, in which quantum chaos has set in, the eigenfunctions are
fully extended.   In this situation the eigenstates  can be considered
as random states with uniformly distributed components with amplitudes
$c_n \approx  1/\sqrt{\mathcal{N}}$ and  random phases. In  this case,
the density matrix can be written as
\begin{equation}
\rho \approx \mathrm{diag}\left(1/\mathcal{N},1/\mathcal{N},\ldots,1/\mathcal{N}\right) + \Omega \ ,
\end{equation}
where  $\Omega$  is  a  $\mathcal{N}\times\mathcal{N}$  zero  diagonal
matrix  with  random complex  matrix  elements  of amplitude  $\approx
1/\mathcal{N}$.

Suppose now that we partition the Hilbert space of the system into two
parts  with  dimensions  $\mathcal{N}_A$  and  $\mathcal{N}_B$,  where
$\mathcal{N}_A\mathcal{N}_B=\mathcal{N}$.  The  reduced density matrix
$\rho_A$ is defined as follows:
\begin{equation}
\rho_A=\mathrm{Tr}_B \rho=
\sum_{n_B} c_{n_A n_B} c^\star_{n_A^\prime n_B} 
|n_A\rangle \langle n_A^\prime|\,,
\end{equation}
where  $|n\rangle =  |n_A n_B\rangle$.   Therefore, in  the integrable
case  the reduced  density matrix  $\rho_A$ is  a $\mathcal{N}_A\times
\mathcal{N}_A$ matrix given by
\begin{equation} \label{eq:red-dens-mat-int}
\rho_A \approx \mathrm{diag}\left(0,0,\ldots,0,1,0,\ldots,0\right) \ .
\end{equation}
In contrast, in the chaotic case 
\begin{equation} \label{eq:red-dens-mat-ch}
\rho_A \approx \mathrm{diag}
\left(1/\mathcal{N}_A,1/\mathcal{N}_A,\ldots,1/\mathcal{N}_A\right) + \Omega_A\ ,
\end{equation}
where $\Omega_A$  is a  zero diagonal matrix  with matrix  elements of
$\mathcal{O}\left(\sqrt{\mathcal{N}_B}/\mathcal{N}\right)$   (sum   of
$\mathcal{N}_B$ terms of order $1/\mathcal{N}$ with random phases).

With  Eqs.~(\ref{eq:red-dens-mat-int})  and (\ref{eq:red-dens-mat-ch})
in  hand, it  is easy  to  calculate the  values for  the reduced  Von
Neumann entropy and the concurrence. We have
\begin{equation} \label{eq:S-ext-values}
S_A \approx \left\{
\begin{array}{ccl}
0 & \mathrm{for} & \mathrm{integrable}\; \mathrm{regime} \\
\log (\mathcal{N}_A) & \mathrm{for} & \mathrm{chaotic}\; \mathrm{regime} \\
\end{array}
\right..
\end{equation}
A better estimate of $S_A$  is obtained by considering the ensemble of
random   states   according   to   the   Haar   measure:   $S_A\approx
\log(\mathcal{N}_A)-\mathcal{N}_A/(2\mathcal{N}_B\log_e             2)$
\cite{random-states}.

For the  case of concurrence  the dimension $\mathcal{N}_A=4$.   If in
Eqs.~(\ref{eq:red-dens-mat-int})-(\ref{eq:red-dens-mat-ch})  
we neglect  the  matrix $\Omega_A$  we obtain
\begin{equation} \label{eq:Cl-ext-values}
c_\lambda \approx \left\{
\begin{array}{ccl}
0 & \mathrm{for} & \mathrm{integrable}\;\mathrm{regime} \\
-1/2 & \mathrm{for} & \mathrm{chaotic}\;\mathrm{regime} \\
\end{array}
\right..
\end{equation}
Thus,  in  both  integrable  and  chaotic  extremes,  we  obtain  from
Eq.~(\ref{eq:C}) that  the concurrence  is zero.

\section{Chaos and entanglement in spin chains}
\label{sec:examples}

In  this  Section,  we  discuss bipartite  and  pairwise  entanglement
measures in two  quantum lattice spin models in  which a transition to
chaos has  been previously found and characterized.   Both models have
been proposed as  suitable model for quantum computers.   For the sake
of  completeness in  this section  we review  the known  properties of
these models, in particular the  onset of quantum chaos.  In parallel,
we   present  new   results   concerning  the   behavior  of   quantum
entanglement.    In  section~\ref{sec:2D-model}   we   focus,  for   a
two-dimensional spin  lattice, on the  behavior of the  concurrence as
quantum chaos  sets in.  These results qualitatively  agree with those
presented    in   section~\ref{sec:1D-model}    for   a    family   of
one-dimensional  spin  models, for  which  we  present  a much  deeper
analysis of  the behavior  of the concurrence  and of the  Von Neumann
entropy across the transition to quantum chaos.

\subsection{Two-dimensional spin lattice}
\label{sec:2D-model}

We consider consists of $L$ spin-$1/2$  particles (qubits) placed on a
two-dimensional square lattice  in the presence  of an external static
magnetic field   directed along $z$.   Nearest-neighbor spins interact
via   Ising coupling with random   strength.   The Hamiltonian of  the
system is
\begin{equation} \label{eq:2D-H}
H = \sum_i \Gamma_i \sigma_i^z + \sum_{i<j} J_{i,j} \sigma_i^x\sigma_j^x \ .
\end{equation}
The operators $\sigma_i$ are the standard Pauli matrices acting on the
$i$-th  qubit.    The  second  sum  in  the   Hamiltonian   runs  over
\emph{nearest-neighbor} spins  and  periodic  boundary  conditions are
considered.  $\Gamma_i$  corresponds to the energy  separation between
the states  of the qubit  $i$.  $J_{i,j}$ is  the interaction strength
between  the   qubits  $i$  and $j$.   The   parameters $\Gamma_i$ and
$J_{i,j}$  are randomly and    uniformly distributed in  the intervals
$[\Delta_0   -   \delta/2,  \Delta_0    + \delta/2]$    and  $[-J,J]$,
respectively.   This Hamiltonian was proposed  as  a model of isolated
quantum computer with hardware imperfections \cite{georgeot00}.

\begin{figure}[!t]
\begin{center}
\includegraphics[scale=0.4]{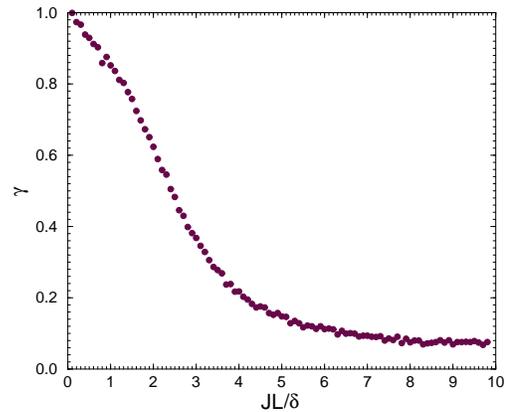}
\caption{\label{fig:2D-gamma} Level statistics parameter $\gamma$ as a
function of  the coupling parameter $J$  (in units of $\delta/L$), for
the model of Eq.~(\ref{eq:2D-H}) for a $3\times 3$ lattice, calculated
from the  energy levels in  the spectral  band centered at $-\Delta_0$
averaged  over      2000 random  realizations     of   $\delta_i$  and
$J_{i,j}$. The  other  parameters are  $\Delta_0  = 1$  and  $\delta =
0.09$.}
\end{center}
\end{figure}

Here we  focus on the  case $\delta,J \ll \Delta_0$, which corresponds
to the situation  where fluctuations induced by lattice  imperfections
are relatively weak.   In  this case, the unperturbed  energy spectrum
($J_{i,j}=0$) of Hamiltonian (\ref{eq:2D-H}) is composed by $L+1$ well
separated bands,  with  inter-band  spacing  $2\Delta_0$.   Each  band
corresponds  to states with a given  number of  spins ``up'' and spins
``down''.  The  highest density of states is  obtained for the central
energy band and therefore we expect  that quantum chaos shows up first
there.  When  interaction is turned  on, a  transition  to chaos takes
place.    A   value for the  chaos  border   $J_{\mathrm c}$  for this
transition      was    given     in  \cite{georgeot00}:    $J_{\mathrm
c}\propto\delta/L$.  This border was corroborated in \cite{benenti01},
where  the emergence of   Fermi-Dirac  thermalization in the   chaotic
regime was studied.  A careful and detailed analysis of the transition
to chaos for this model and its dependence  on the size of the lattice
has been taken  in \cite{georgeot00,benenti01}.  For the sake
of comparison with  the   behavior of the  entanglement  measures we
repeat  some   of  these previous    results.  We  consider  a  square
$3\times3$   lattice.  We note that for    lattices composed of an odd
number of qubits there  is not a central  energy band but instead, two
central  bands  centered at  $\pm\Delta_0$.   In what follows, we will
consider the states from the central band centered at $-\Delta_0$.

We have   numerically   diagonalized Hamiltonian   (\ref{eq:2D-H}) for
different values of the interaction strength.  To study the transition
to chaos we  have  obtained the  spectral  statistics in terms of  the
nearest neighbor spacing distribution  $P(s)$ as well as the structure
of the eigenfunctions in terms of  the participation number $\xi$.  We
have   restricted our calculations    to the energies  and eigenstates
encountered  in the central  negative band.  For weak interactions the
energy domain of  this band is  clearly visible  and we keep  the same
domain  even  for stronger    interaction   where the  band  structure
disappears.

In Fig.~\ref{fig:2D-gamma} the parameter $\gamma$ as a function of the
interaction strength is shown for $\Delta_0  = 1$ and $\delta = 0.09$.
When   the strength  of   the interaction   $J$  increases the  $P(s)$
distribution  smoothly changes from  Poisson ($\gamma=1$)  towards GOE
($\gamma=0$).  Thus, increasing  the interaction between the  qubits a
transition to  quantum  chaos occurs.   In Fig.~\ref{fig:2D-gamma}  we
observe that the crossover from integrability  to chaos takes place in
the interval between  $JL/\delta \approx 1$ and $JL/\delta\approx  5$.

\begin{figure}[!t]
\begin{center}
\includegraphics[scale=0.4]{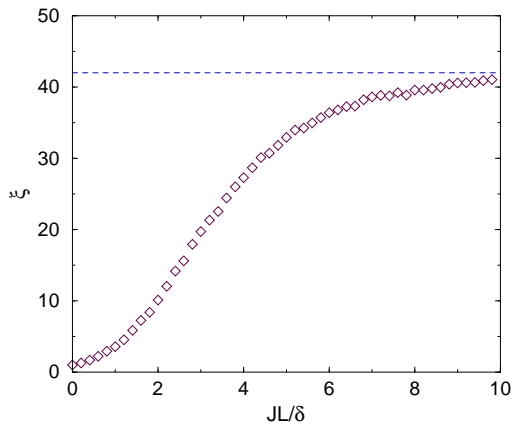}
\caption{\label{fig:2D-PN}
Participation number $\xi$ as a  function of $JL/\delta$ for the model
of Eq.~\ref{eq:2D-H})   for  a $3 \times  3$  lattice,  with parameter
values as  in Fig.~\ref{fig:2D-gamma}.  The  PN  was obtained from the
central eigenfunction   of the band  centered at  $-\Delta_0$ averaged
over   2000  random realizations.    The  dashed  line  corresponds to
$N_b/3$.}
\end{center}
\end{figure}

At  the   same time,  the  eigenfunctions   start  to mix.   For  weak
interactions the eigenfunctions  are strongly localized: The number of
components in the basis of the unperturbed Hamiltonian is of the order
of one.  At  strong interactions ($JL/\delta\gg 1$) the eigenfunctions
are extended,  having a large   number of non  negligible  components.
This mixing of the eigenfunctions is shown in Fig.~\ref{fig:2D-PN} in
terms   of    the   PN for   the     same  parameter  values   as  in
Fig.~\ref{fig:2D-gamma}.  In Fig.~\ref{fig:2D-PN} we see that the PN
smoothly changes from $1$ (localized regime) to  its upper bound value
of $N_b/3$ (chaotic regime), where  $N_b$ corresponds to the number of
eigenfunctions with energies in the central negative  band. As we have
discussed, the factor  of $1/3$  arises due to  the symmetries  of the
chaotic Hamiltonian that are described by the GOE.

\begin{figure}[!t]
\begin{center}
\includegraphics[scale=0.4]{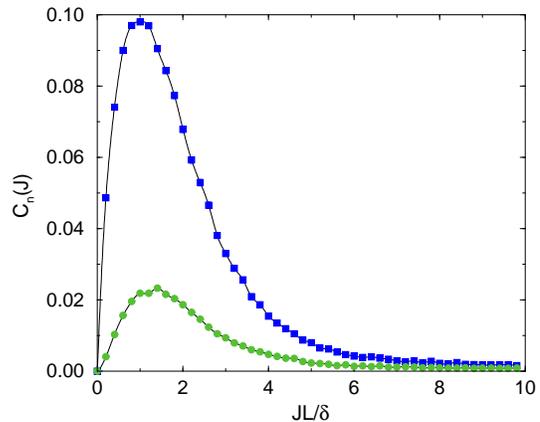}
\caption{\label{fig:2D-concurrence}
Concurrence $C_n$ as a function of  the coupling parameter $JL/\delta$
for the model of Eq.~(\ref{eq:2D-H}) for a  $3 \times 3$ lattice, with
parameter values  as in  Fig.~\ref{fig:2D-gamma}.  $C_n$ was  obtained
from the  central  eigenfunction of the central  negative  band of the
spectrum   and  averaged over all  possible    pairs of:  $a$) nearest
neighbors qubits ($n=1$) and  $b$)  next to nearest  neighbors  qubits
($n=2$).}
\end{center}
\end{figure}

We  now turn  our attention  to  the entanglement  measures.  We  have
calculated the concurrence between nearest ($C_1$) and next to nearest
($C_2$) neighbor qubits.  For this purpose we have drawn $2000$ random
realizations of $\delta_i$  and $J_{i,j}$ and diagonalized Hamiltonian
(\ref{eq:2D-H}).   Using   for  each  realization   only  the  central
eigenfunction of the central negative band we have calculated the mean
concurrence  averaged  over all  possible  nearest  neighbor pairs  of
qubits.  In Fig.~\ref{fig:2D-concurrence}  we show $C_1$ (squares) and
$C_2$  (circles),  averaged over  all  the  random  realizations as  a
function  of the strength  of the  interaction.  In  the basis  of the
unperturbed Hamiltonian  the concurrence  is strictly zero.   For very
weak interactions  ($JL/\delta\ll 1$), the  concurrence remains small.
At   the  other  extreme,   when  the   interaction  is   very  strong
($JL/\delta\gg 1$)  and quantum chaos  has set in, the  concurrence is
also  small, as  expected.  Quite  interestingly, the  maximum  of the
$C_1$ concurrence is for $JL/\delta \approx 1$, that is, in the region
in  which the  crossover  from integrability  to  quantum chaos  takes
place.   In  Fig.~\ref{fig:2D-concurrence}  we  can also  compare  the
behavior  of $C_1$ with  that of  $C_2$.  The  concurrence of  next to
nearest  neighbor qubits is  noticeable smaller  than that  of nearest
neighbor qubits.  This  is not surprising as the  Ising interaction in
Eq.~(\ref{eq:2D-H}) couples only  nearest neighbor qubits.  Therefore,
one  should  expect  that  quantum  correlations  between  qubits  are
stronger  for qubits  that  are  close than  for  those farther  away.
However,  we find  that $C_2$  is not  negligible everywhere  over the
domain  of $J$  investigated, except  for the  integrable  and chaotic
extremes,  where also  $C_1$ goes  to zero.   Similarly to  $C_1$, the
concurrence $C_2$ has its maximum value for $JL/\delta \approx 1$.  It
is interesting to point out that, similarly to what happens in the QPT
in  the Ising model  \cite{fazio02,osborne02}, the  concurrences $C_1$
and $C_2$  exhibit their  maximum values close  to the value  at which
quantum chaos  sets in.  However,  besides this similarity,  there are
other aspects in which the  entanglement at the onset of chaos behaves
in   a  different   manner  than   for  the   case  of   a   QPT.   In
section~\ref{sec:1D-model},   the  qualitative   differences   of  the
behavior of  entanglement in a QPT  an for the onset  of quantum chaos
that we observe will be discussed.

\begin{figure}[!t]
\begin{center}
\includegraphics[scale=0.45]{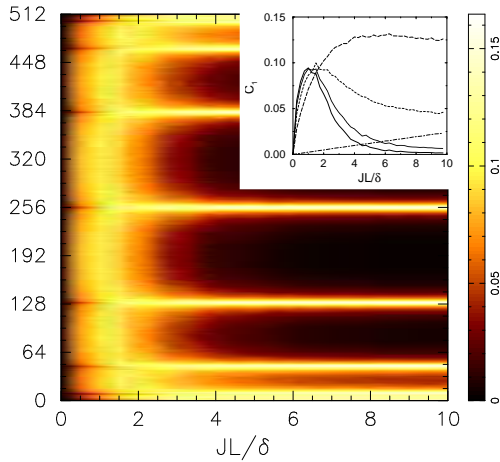}
\caption{\label{fig:2D-C-bands}
Color   density  map for  the   concurrence $C_1$ calculated  for each
individual eigenfunction (index  in the vertical  axis), averaged over
200 random realizations. $C_1$is plotted as a function of the coupling
parameter $JL/\delta$ for the   model of Eq.~(\ref{eq:2D-H}) for  a $3
\times    3$       lattice,  with       parameter   values     as   in
Fig.~\ref{fig:2D-gamma}.  In the inset, we   show $C_1$ averaged  over
the central  $1/3$  of the eigenstates of   each individual  band.  The
different curves correspond to the bands for states with: $4$ (solid),
$3$   (dotted), $2$ (dashed),   $1$ (long-dashed) and $0$ (dot-dashed)
spins ``up''.}
\end{center}
\end{figure}

As we have discussed  in section~\ref{sec:chaos}, the onset of quantum
chaos  occurs when  the  typical interaction  matrix elements  between
directly  coupled states  becomes of  the  order of  their mean  level
spacing.  Thus, the onset of  quantum chaos is expected to be observed
first  at the  spectral  energies  at which  the  density of  directly
coupled states is  larger.  For the models discussed  in this and next
sections this happens  at the center of the  spectrum, \emph{i.e.}, in
the  central bands.   Consequently,  we focus  our  discussion on  the
eigenstates  corresponding   to  the  central  energy   bands.  It  is
worthwhile mentioning that our choice is different from the studies of
QPT for which  the transition is a property of  the ground state.  For
the   case   of  a   transition   to   quantum   chaos  we   show   in
Fig.~\ref{fig:2D-C-bands}  the value  of $C_1$  as a  function  of the
coupling parameter $J$, calculated  for each eigenstates of a $3\times
3$ lattice.  The band structure can be clearly seen in the behavior of
the concurrence.  Inside each band, the concurrence grows from zero at
$J\sim 0$ to  a maximum close to $JL/\delta=1$,  after which it decays
again to zero as $J$ is  further increased. On the other hand, for the
states at the  border of the bands, including the  ground and the most
energetic states, $C_1$ behaves quite  differently. This is due to the
fact that these states do not mix significantly with the other states,
as confirmed by our  numerical data computing the participation number
(data not shown).  In  the inset of Fig.~\ref{fig:2D-C-bands}, we show
$C_1$, averaged over the central  eigenstates of each band. We observe
that, for all bands, $C_1$ reaches  its maximum value for $J$ close to
$\delta/L$.  The fact that for  the bands farther away from the center
of the spectrum  $C_1$ decays slower to zero as $J$  is increased is a
finite size  effect that should disappear at  the thermodynamic limit.
However, for the  ground state $C_1$ grows linearly.   While we do not
discard  that   at  the  thermodynamic  limit   even  the  concurrence
calculated from the  ground state will behave in  a similar fashion as
for  any   other  typical  state,   Fig.~\ref{fig:2D-C-bands}  clearly
indicates that for a finite system this is not the case.

\subsection{One-dimensional spin chain}
\label{sec:1D-model}

In this section  we  discuss the behavior   of bipartite and  pairwise
entanglement in a family of one-dimensional spin $1/2$ chains.  Due to
its  lower dimensionality these models  will  allow us to characterize
the behavior of  the entanglement across the  transition to chaos in a
deeper fashion  than for the  previous model.  We  shall find the same
behavior for the   concurrence than before.  Nevertheless,  with these
models we are able  to analyze its  dependence on: the distance in the
lattice between the partners, the range of the interaction and the size
of the chain.  Moreover, for  one member of  this family of models the
chaos  border does not coincide with  the  delocalization border. This
will   give us   the  possibility to  compare    the behavior  of  the
concurrence in a regime of weak and of hard chaos.

\subsubsection{Definition of the model.}

We consider  a   system consisting on  a   linear chain of   $L$ $1/2$
interacting spins,  subjected  to a static  transverse  magnetic field
(along $z$)  and to a circularly  polarized magnetic field rotating in
the   $(x,y)$     plane  with  frequency     $\nu$,   ${\vec  B}(t)  =
(B^\perp\cos(\nu t+\varphi),-B^\perp\sin(\nu t+\varphi), B^z)$
\cite{berman01}.  In  the coordinate system, which  rotates around the
$z$ axis with  frequency $\nu$, the Hamiltonian of  this system can be
written as
\begin{equation} \label{eq:1D-H-rot}
\begin{array}{ll}
H   =  &-\frac{1}{2}\sum\limits_{k=1}^{L}  \{\delta_k\sigma^z_k+
\Omega(\cos\varphi ~ \sigma^x_k-\sin\varphi ~ \sigma^y_k)\}  \\  & +
\frac{1}{2}\sum\limits_{k=1}^{L-1}J_{k,k+1}\sigma^z_k \sigma^z_{k+1},
\end{array}
\end{equation}
$\delta_k=\omega_k-\nu$  where  $\omega_k$  is  the frequency  of  the
precession of the $k$-th spin  in the $B^z$ field. $\Omega$ stands for
the Rabi frequency corresponding to the rotating field and $J_{k,k+1}$
denotes the  strength of the  Ising interaction between the  spins $k$
and $k+1$.  The operators  $\sigma_k$ are the standard Pauli operators
acting  on  the $k$-th  spin.   In the  following,  we  will take  for
simplicity $\varphi = \pi/2$ and  consider that the static field $B^z$
has  a constant gradient  $a$ along  the chain  such that  $\delta_k =
ak$. Thus, the Hamiltonian takes the form

\begin{equation} \label{eq:1D-H-dyn}
H = \frac{1}{2}\sum_{k=1}^L \left(-\delta_k\sigma_k^z + \Omega
\sigma_k^y\right) - \frac{1}{2}\sum_{k=1}^{L-1}J_{k,k+1}\sigma_k^z
\sigma_{k+1}^z \ .
\end{equation}

We assume that for all $k$ the inequality $\Omega \gg \delta_k$ holds.
Open boundary  conditions are  taken.  In \cite{berman01},  this model
was proposed  as a possible candidate for  experimental realization of
quantum  computation.   The  gradient  of magnetic  field  provides  a
labeling of  qubits in  terms of their  Larmor frequencies.   Thus, it
allows for a way to address each qubit separately.

It   is   worthwhile    mentioning   that,   besides   the   different
dimensionality, there  is a more striking difference  between this and
the  previous model:  The  existence  of a  constant  gradient in  the
magnetic field gives rise to a $L$-independent threshold for the onset
of (weak)  chaos.  In \cite{berman01} the transition  to quantum chaos
and   its   implications  to   quantum   computation  were   explored.
Here we want to discuss the behavior of entanglement in this model.

In order to apply the approach discussed in section~\ref{sec:chaos} it
is  convenient  to represent  Hamiltonian  (\ref{eq:1D-H-dyn}) in  the
basis  in which  it is  diagonal for  non-interacting spins.   In this
so-called  effective field  representation,  the one-body  unperturbed
Hamiltonian $H_0$ takes the form
\begin{equation} \label{eq:1D-H0}
H_0 = \frac{1}{2}\sum_{k=1}^L \sqrt{\delta_k^2 + \Omega^2} \ \sigma_k^z
\end{equation}
and the interaction Hamiltonian $V$  can be written as
\mbox{$V = V_{\sf diag} + V_{\sf band} + V_{\sf off}$}, where
\begin{eqnarray} \label{eq:1D-V-NN}
V_{\sf diag} & = & -\frac{1}{2}\sum_{k=1}^{L-1}J_{k,k+1}b_kb_{k+1}
\sigma_k^z\sigma_{k+1}^z,\nonumber \\
V_{\sf band} & = & -\frac{1}{2}\sum_{k=1}^{L-1}J_{k,k+1}a_ka_{k+1}
\sigma_k^y\sigma_{k+1}^y, \\
V_{\sf off}  & = & \frac{1}{2}\sum_{k=1}^{L-1}J_{k,k+1}\left(a_kb_{k+1}
\sigma_k^y\sigma_{k+1}^z + a_{k+1}b_k\sigma_k^z\sigma_{k+1}^y\right),\nonumber
\end{eqnarray}
with
\begin{equation} \label{eq:1D-V-param}
a_k = \frac{\Omega}{\sqrt{\delta_k^2 + \Omega^2}} \quad , \quad 
b_k = \frac{-\delta_k}{\sqrt{\delta_k^2 + \Omega^2}} \ .
\end{equation}
As before, the quantities $J_{k,k+1}$ stand for the Ising interactions
between nearest neighbor spins. In  what follows, we will consider the
interactions  to  be completely  random,  {\it i.e.}  $J_{k,k+1}=J\xi$,
where $\xi$ is  a random number uniformly distributed  in the interval
$[-1,1]$.  This  model is  known  as the  NN  model  from its  nearest
neighbor character.

\begin{figure}[!t]
\begin{center}
\includegraphics[scale=0.4]{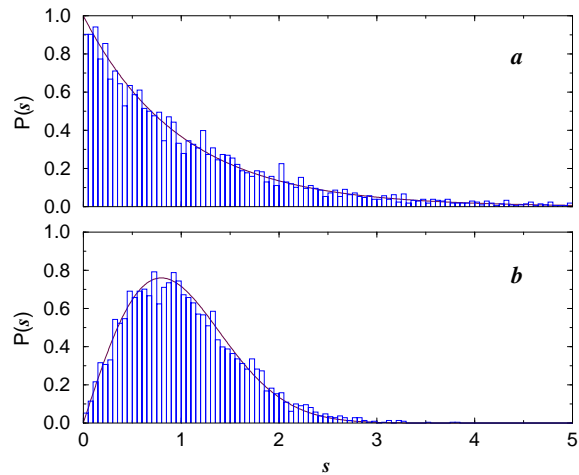}
\caption{\label{fig:1D-pds}
Nearest  neighbor level spacing  distribution $P(s)$ for  the AA model
($l_c  = L-1$)of Eq.~(\ref{eq:1D-V}) for  a chain  of $12$ qubits and:
$a$)  $J/J_{\mathrm c}  = 0.35$   and   $b$) $J/J_{\mathrm  c} =  15$,
calculated from the energy levels in the  central band of the spectrum
and averaged over 10 random realizations.   The solid lines correspond
to the   Poisson     ($a$) and  Wigner    surmise  ($b$)   theoretical
distributions.}
\end{center}
\end{figure}

As  in the two-dimensional  model (\ref{eq:2D-H}), for the unperturbed
($J=0$) case the spectrum  possesses a band  structure.  Each band  is
characterized  by  a  constant number   $n$  of qubits   in the  state
$\ket{0}$ and $L-n$ qubits in the state  $\ket{1}$.  When $L$ is even,
a central  band  (around zero) exists. It  consist   of the many-qubit
states with  $L/2$ spins ``up'' and $L/2$  spins ``down''.  The number
of these states is given by
\begin{equation} \label{eq:Nb}
N_b = \frac{L!}{\left(L/2\right)!\left(L/2\right)!} \ .
\end{equation}
As discussed at the end of previous section, we will only consider the
energy levels and energy eigenstates corresponding to the central band
of the spectrum.

\begin{figure}[!t]
\begin{center}
\includegraphics[scale=0.4]{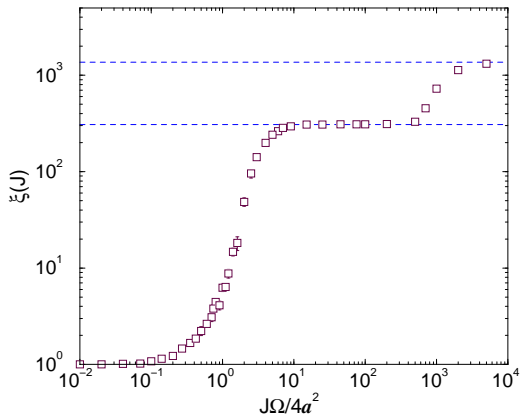}
\caption{\label{fig:1D-PN}
Participation number $\xi$ as a function of the coupling parameter $J$
for the AA model  of Eq.~(\ref{eq:1D-V}), for a  chain of size $L=12$.
The PN is averaged over all eigenfunctions in  the central band of the
spectrum and over 10 random realizations.  The dashed horizontal lines
correspond to $N_b/3$ and  to $N/3$ where $N=2^L$  is the dimension of
the Hilbert space.}
\end{center}
\end{figure}

When $J > 0$ the potential term $V$  mixes the states inside each band
and among different  bands: In the  basis of $H_0$,  $V_{\sf diag}$ is
diagonal.  Instead, $V_{\sf band}$ couples  states which are either in
the same   band or in next  to  nearest bands.  $V_{\sf  off}$ couples
states  which  are in nearest neighbor   bands.  The  mixing of energy
bands  triggers the  transition  to   chaos.  For  a relatively   weak
interaction the   eigenstates (in the  basis  of $H_0$) are localized,
while for stronger interaction the  number of components significantly
different from zero increases.  The transition from strongly localized
to   extended   states occurs very  fast  with   the increase   of the
interaction and  sets in when the strength  of the typical interaction
is of the  order of the mean  level  spacing between  directly coupled
many-body states \cite{many-bodies}.  In \cite{berman01} the value for
the   delocalization border was   found  to be $J_{\mathrm{c}} \approx
4a^2/\Omega$.

However, as it was shown in \cite{berman01}, the  NN model is peculiar
in the following sense:  The  delocalization border does  not coincide
with  the  chaos border.  Increasing  the  strength of the interaction
$J$, the system  goes from  a  regular regime  to a \emph{weak  chaos}
regime where the     eigenfunctions  are delocalized but   the   level
statistics is  not  yet described by Random    Matrix Theory.  If  the
interaction   is further increased the   bands overlap  and the system
enters a regime of  \emph{strong chaos}.  This peculiarity is  removed
if the range of the interaction is larger than nearest neighbor.

Here we will  consider a range of  the interaction $l_c$ from $1$ (for
the  NN model) up to  $L-1$.  The interaction term  $V$ keeps the same
structure as in Eq.~(\ref{eq:1D-V-NN}) but the different terms are now
\begin{eqnarray} \label{eq:1D-V}
V_{\sf diag} & = & -\frac{1}{2}\sum_{j=1}^{L-1} \ \sum_{k=j+1}^{j+l_c\le L}
J_{jk}b_jb_k\sigma_j^z\sigma_k^z,\nonumber \\
V_{\sf band} & = & -\frac{1}{2}\sum_{j=1}^{L-1} \ \sum_{k=j+1}^{j+l_c\le L}
J_{jk}a_ja_k\sigma_j^y\sigma_k^y, \\
V_{\sf off}  & = & \frac{1}{2}\sum_{j=1}^{L-1} \ \sum_{k=j+1}^{j+l_c\le L}
J_{jk}\left(a_jb_k\sigma_j^y\sigma_k^z + a_kb_j\sigma_j^z\sigma_k^y
\right)\nonumber \ .
\end{eqnarray}
For $l_c = L-1$ this model is known as the AA (All to All) model as in
this  case all qubits  are allowed  to interact  with each  others. In
contrast with  the NN model ($l_c=1$),  if $l_c > 1$  the chaos border
occurs at the same  value $J_{\mathrm{cr}} \approx 4a^2/\Omega$ as the
delocalization  border  \cite{berman01}.

\subsubsection{The onset of quantum chaos.}

In  Fig.~\ref{fig:1D-pds}  the nearest  level  spacing distribution is
shown  for  the AA  model   for a chain   of  $12$ qubits. The  $P(s)$
distribution  was obtained  from the  energy  levels contained  in the
central band and averaged over $10$ different random realizations.  In
panel $a$, the case of weak interaction  ($J/J_{\mathrm c} = 0.35$) is
shown.   It is in good agreement  with the Poisson distribution (solid
line), as expected in the  integrable regime.  In contrast, panel  $b$
shows the situation corresponding to strong coupling ($J/J_{\mathrm c}
= 15$) in  which  the $P(s)$ distribution  follows the  Wigner surmise
expected for a chaotic system with GOE statistics.   In panel $b$, the
level   repulsion   effect is evident.     Thus, when  the interaction
strength $J > J_{\mathrm  c}$,  the spectral level  statistics changes
from Poisson  to GOE showing  that a  transition to quantum  chaos  is
happening.

\begin{figure}[!t]
\begin{center}
\includegraphics[scale=0.4]{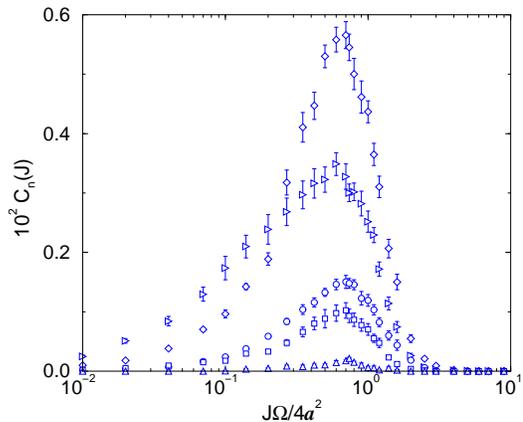}
\caption{\label{fig:C_vs_distance}
Concurrence as a function of the coupling parameter  $J$ for the model
of  Eq.~(\ref{eq:1D-V}) with interaction  range $l_c=5$ for a chain of
$10$ qubits.  For each eigenfunction $C_n$ was obtained as the average
concurrence between all   possible  pairs of  qubits separated  by   a
distance $n$: $n=1$ for nearest neighbors  pairs (diamonds), $n=2$ for
next to  nearest  neighbors pairs (right-triangles), $n=3$  (circles),
$n=4$  (squares)    and  $n=5$ (up-triangles).  The     plotted values
correspond  to $C_n$ averaged over  all the eigenstates of the central
band and over 30 random realizations.}
\end{center}
\end{figure}

Simultaneously,  a  localization-delocalization transition  occurs for
the eigenstates in the central band.   This transition takes place for
any value  of $l_c$.  However, for the  NN model, the PN does saturate
at a value which  is lower than $N_b/3$ corresponding  to the  case of
Gaussian fluctuations.  In  Fig.~\ref{fig:1D-PN}, the PN is shown  for
the AA model for a chain composed of $12$ qubits.  The PN (squares) is
averaged over all the  eigenstates in the  central band and  over $30$
different random realizations.  For weak interactions, the eigenstates
are  effectively   localized   ($\xi\approx  1$).   The   PN increases
monotonously with the interaction until it  reaches the value $N_b/3 =
308$ (lower dashed line).  A complete mixing of different bands occurs
for much stronger interactions ($J/J_{\mathrm c} \approx 1000$).  This
is seen   in Fig.~\ref{fig:1D-PN}, where    $\xi$ increases again  and
reaches its upper bound value corresponding $N/3$, namely to one third
of the dimension of the whole Hilbert space, as expected from RMT.

The model  of Eq.~(\ref{eq:1D-V}) shows a clear  transition to quantum
chaos  in  which  both  energy  levels and  eigenstates  change  their
character.  Now we  turn  our  attention to  the  behavior of  quantum
entanglement.

\begin{figure}[!t]
\begin{center}
\includegraphics[scale=0.4]{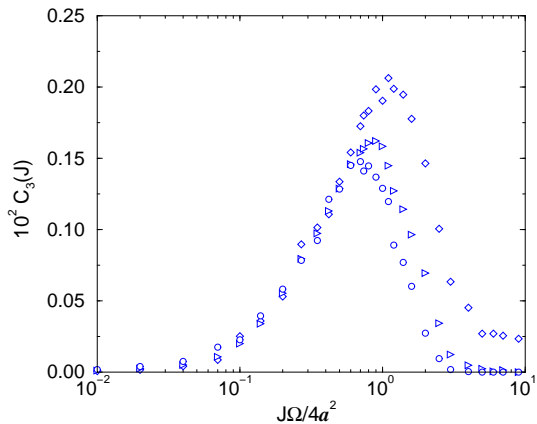}
\caption{\label{fig:C_vs_interaction}  Mean concurrence  $C_3$ between
qubits at  a distance $n=3$,  averaged over all the  eigenfunctions in
the central band of the  spectrum and over 30 random realizations. The
mean concurrence  is plotted as  a function of the  coupling parameter
$J$ for  the model of Eq.~(\ref{eq:1D-V})  in a chain  of $10$ qubits.
We compare $C_3$ for different  range of the interaction: The NN model
corresponding to $l_c=1$  (diamonds), $l_c=2$ (triangles), and $l_c=3$
(circles).}
\end{center}
\end{figure}

\subsubsection{The concurrence: Sharing of entanglement.}

We   have calculated  the  mean  concurrence   averaged  over all  the
eigenstates in the central band  as a function of  the strength $J$ of
the interaction.  In Fig.~\ref{fig:C_vs_distance} the mean concurrence
is shown  for the model of  Eq.~(\ref{eq:1D-V}) for  a chain of $L=10$
qubits, with an interaction  that couples $L/2$ neighbor  qubits, {\it
i.e.}, $l_c  = 5$.    The diamond symbols    correspond  to the   mean
concurrence    $C_1$ between nearest  neighbor  qubits   .  As it  was
discussed in section~\ref{sec:borders}, we observe that $C_1$ is close
to zero in both   extremes of chaos  and of  integrability.  Moreover,
similarly to  what  we  have  observed for  the  two-dimensional model
(\ref{eq:2D-H}),  in between the integrable  and  the chaotic extremes
the concurrence increases and its maximum  value is close to the value
for the  chaos border.  Despite the  fact  that we have  observed this
behavior  of concurrence for just two   different models we conjecture
that    it is    generic      for  transitions    to    chaos.      In
Fig.~\ref{fig:C_vs_distance} the  mean concurrence  averaged over  all
qubits at further distances, $C_2$ (right-triangles), $C_3$ (circles),
$C_4$ (squares) and $C_5$ (up-triangles) are also shown.  The behavior
of $C_1$ and $C_2$ is similar to that observed for the two-dimensional
model.  We see that the  concurrence $C_n$ decreases with the distance
$n$, except for weak interactions for  which $C_2 > C_1$.  This latter
is a peculiarity of this model.
It  is  interesting  to  notice  that the  coupling  strength  $J_{\rm
max}(n)$ at which  the concurrence $C_n$ takes its  maximum value does
not change  significantly with $n$.

Let us now consider the  following question: for a given distance $n$,
how does $C_n$  varies if the range of  the interaction increases?  In
Fig.~\ref{fig:C_vs_interaction} the mean  concurrence $C_3$ is plotted
for  different ranges  $l_c$ of  the interaction:  $l_c=1$ (diamonds),
$l_c=2$  (triangles)  and  $l_c=3$  (circles).  To  this  purpose,  we
computed $C_3$ for  all the eigenfunctions in the  central band of the
spectrum of  a chain of size  $L=10$ and averaged  over $30$ different
random realizations.   From Fig.~\ref{fig:C_vs_interaction} we observe
that as  the range of  the interaction increases the  mean concurrence
$C_3$  decreases.  The  same conclusions  were also  obtained  for the
behavior  $C_n$ with  $n\ne3$  (data  not shown).   This  fact can  be
understood from the pairwise  character of the concurrence.  Since the
amount of entanglement between one  definite qubit and the rest of the
system is bounded, this finite amount of entanglement has to be shared
between all possible  partners.  When the range of  the interaction is
enlarged, it  becomes easier for  each qubit to become  entangled with
more  qubits in  the chain.   As  a result  the pairwise  entanglement
between one  single qubit and  the rest of  the chain is  shared among
more partners.  Therefore, the average entanglement shared between two
qubits decreases.  This argument is valid  if a change in the range of
the  interaction does  not significantly  change the  total  amount of
bipartite entanglement that is shared between a single qubit and the
rest of the chain.

\begin{figure}[!t]
\begin{center}
\includegraphics[scale=0.4]{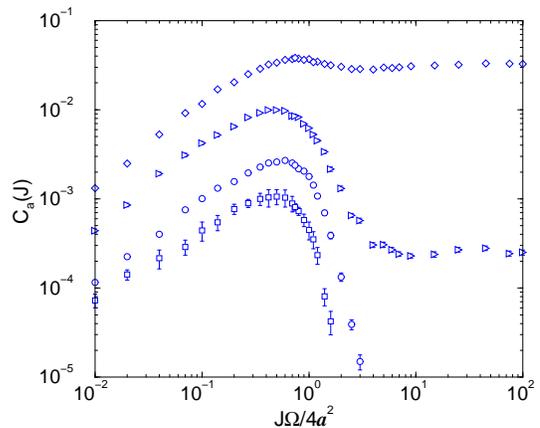}
\caption{\label{fig:C_vs_size}  Mean  concurrence  averaged  over  all
possible pairs of  qubits for all the eigenstates  in the central band
of the  spectrum and over 30  random realizations for the  AA model of
Eq.~(\ref{eq:1D-V})  ($l_c=L-1$)   as  a  function   of  the  coupling
parameter $J$  and for different sizes  of the chain  of qubits: $L=6$
(diamonds), $L=8$ (triangles), $L=10$ (circles), $L=12$ (squares).}
\end{center}
\end{figure}

Finally, we have studied the behavior of the concurrence as a function
of the size  of   the system.  In Fig.~\ref{fig:C_vs_size}  the   mean
concurrence $C_a$ obtained from all  possible pairs of qubits is shown
for  the AA model for different  sizes of the chain: $L=6$ (diamonds),
$L=8$ (triangles), $L=10$ (circles),  $L=12$ (squares).  In this case,
instead of measuring the concurrence  $C_n$ for some  value of $n$, we
have measured the concurrence  $C_a$  as the mean concurrence  between
all possible pairs  of qubits in the  chain.  This is because, for the
AA model, the concept  of distance turns  out to be meaningless, since
the strength of the interaction between  two qubits does not depend on
their  distance.  The behavior  of $C_a$ as a function  of the size of
the system can be well  understood in terms  of the same argument used
to explain Fig.~\ref{fig:C_vs_interaction}.   When the system size  is
increased, the number of possible  partners  with which a given  qubit
can be entangled also increases.  Therefore, the pairwise entanglement
decreases, in agreement with the data of Fig.~\ref{fig:C_vs_size}.  We
note that,  for small  system sizes  ($L=6,8$),  the concurrence $C_a$
does not  go to zero  at the  chaotic  side  of the transition.   This
finite-size effect disappears already for $L=10,12$. Similar
results were obtained for the other models with different range of the
interaction.

\begin{figure}[!t]
\begin{center}
\includegraphics[scale=0.4]{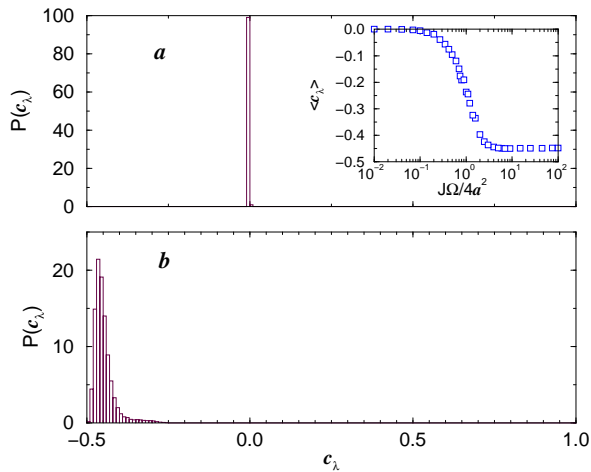}
\caption{\label{fig:cl} Normalized  distribution $P(c_\lambda)$ of the
quantity $c_\lambda = \lambda_1 -  \lambda_2 - \lambda_3 -  \lambda_4$
of Eq.~(\ref{eq:C}) for the AA  model Eq.~(\ref{eq:1D-V}) with  $L=12$
and coupling: $a$) $J/J_{\mathrm c} =  0.01$ and $b$) $J/J_{\mathrm c}
=  100.0$.   In the  inset the behavior  of the  first   moment of the
distribution  $<c_\lambda>$ is  shown as a   function of the  coupling
parameter $J$.}
\end{center}
\end{figure}

The  results  presented in  this  section  show  that the  concurrence
maximizes for  values of  $J$ which  are close to  those at  which the
transition to chaos occurs. As  discussed in the previous section, the
same behavior  for the maximum  of concurrence has also  been observed
for  quantum phase  transitions occurring  in integrable  models (see,
{\it e.g.}, \cite{fazio02,osborne02} for  a study on the Ising chain).
However, as it can  be seen in Fig.~\ref{fig:C_vs_interaction} for the
transition to chaos, the concurrence  approaches zero when the size of
the  system  increases,  in  contrast  to  what  is  observed  in  QPT
\cite{fazio02}, where  $C_n \rightarrow 0$  for all $n>1$  but remains
finite  for  $n=1$.  In  these  studies  a  critical scaling  for  the
derivative of  the concurrence  was obtained.  On  one hand,  the fact
that for  the onset of  quantum chaos the concurrence  diminishes when
the  system approaches  the thermodynamic  limit makes  a  finite size
scaling analysis rather difficult,  as numerical errors become soon of
the same  order of the measure itself.   On the other hand,  it is not
clear whether the transition to chaos in the models we have considered
becomes sharp  at the thermodynamic  limit.  This means that  to prove
the existence of  a critical point for the  transition remains an open
problem.  In  order words, we  lack for a  critical point at  which to
perform  a  scaling  analysis.   We  have  nevertheless  analyzed  the
dependence  of $\mathrm{d}C/\mathrm{d}J$  on the  system  size without
having  found any  clear indication  of a  scaling behavior  (data not
shown).

\begin{figure}[!t]
\begin{center}
\includegraphics[scale=0.4]{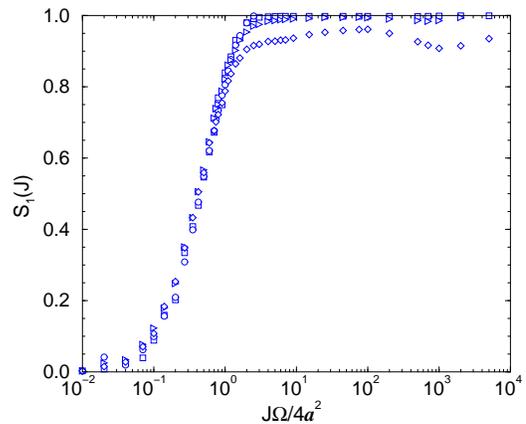}
\caption{\label{fig:S1}  Von Neumann entropy  $S_1$ between  one qubit
and the rest of the system; $S_1$ is averaged over all qubits and over
all eigenstates in the central band of the spectrum and over 30 random
realizations for the AA model  of Eq.~(\ref{eq:1D-V}) as a function of
the coupling  parameter $J$  and for different  sizes of the  chain of
qubits: $L=6$ (diamonds),  $L=8$ (triangles), $L=10$ (circles), $L=12$
(squares).}
\end{center}
\end{figure}

It is interesting  to study the different character  of mixed pairwise
entanglement in  the integrable and  chaotic sides of  the transition.
In Sec.~\ref{sec:borders},  we gave simple arguments  that explain the
different structure of the reduced density matrix $\tilde{\rho}$ for a
mixed bipartite state  in the regimes of integrability  and chaos.  In
the integrable region, due to the localized nature of the eigenstates,
$\tilde{\rho}$ is  essentially diagonal  with only one  matrix element
significantly different from zero.  On  the other hand, in the chaotic
region,  $\tilde{\rho}$ is  almost  diagonal with  matrix elements  of
comparable magnitude along the diagonal.  Both cases give a very small
(or zero) concurrence.  However, while  in the integrable case this is
due to  the fact that  the two-qubit subsystem under  investigation is
essentially  in  a separable  pure  state,  in  the chaotic  case  the
pairwise  entanglement is  zero due  to  the random  structure of  the
eigenfunctions of  the whole $L$-spin  system.  As a  consequence, the
two-qubit reduced density  matrix is essentially diagonal.  Therefore,
in the chaotic regime, the interaction with the rest of the system mimics
a decoherence process for the two-qubit subsystem.

The  different  origin of  the  very  small  value of  concurrence  is
illustrated by  the distribution of  the $c_\lambda$'s (we  remind the
reader  that  the  concurrence  is  defined  as  the  maximum  between
$c_\lambda$ and $0$, see Eq.~(\ref{eq:C})).  In Fig.~\ref{fig:cl}, the
probability  distribution   $P(c_\lambda)$  is  shown   in:  $a$)  the
integrable regime and $b$) the chaotic  regime, for the AA model and a
chain of size  $L=12$. Clearly, in both cases  the probability to find
$c_\lambda  > 0$  is very  small. Therefore,  the concurrence  is very
small in  both cases.   However, the distributions  $P(c_\lambda)$ are
quite   different.   We   note  that   numerical  results   about  the
distribution  $P(c_\lambda)$ in  a  different model  of quantum  chaos
where presented  in \cite{caves}.  In the  Inset of Fig.~\ref{fig:cl},
one can see that the first moment of $P(c_\lambda)$ changes across the
chaos  border.   It  is  an  interesting open  problem  to  obtain  an
analytical   form  of  $P(c_\lambda)$   for  integrable   and  chaotic
situations.   The possibility  to use  this distribution  to  mark the
transition to chaos also deserves more investigation.

\begin{figure}[!t]
\begin{center}
\includegraphics[scale=0.4]{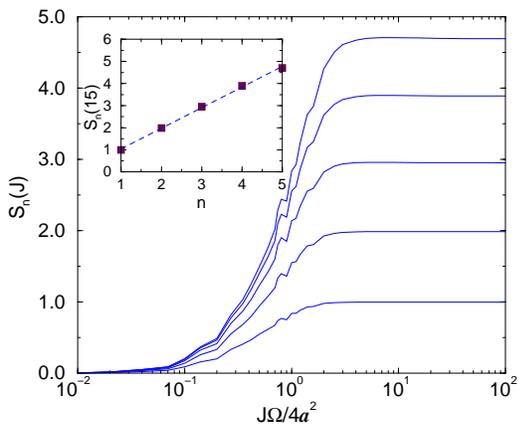}
\caption{\label{fig:S_vs_L} Von Neumann entropy $S_n$ between left and
right  blocks of sizes  $n$ and  $L-n$ qubits,  respectively, averaged
over all eigenfunctions  in the central band of  the spectrum and over
10  random  realizations. The  data  correspond  to  the AA  model  of
Eq.~(\ref{eq:1D-V}) for a  chain of size $L=12$.  From  bottom to top,
the  different  curves  are  from   $n=1$  to  $n=5$.   In  the  inset
$S_n(J/J_{\mathrm c}=15)$ is  plotted as a  function of $n$. The  dashed line
corresponds to a linear fit with a slope of $0.93 \pm 0.02$.}
\end{center}
\end{figure}

\subsubsection{The Von Neumann entropy.}

We now  turn our attention  to the behavior of  bipartite entanglement
measured in terms  of the Von Neumann entropy.   First we consider the
mean Von  Neumann entropy  $S_1$ of  each qubit with  the rest  of the
qubits in  the chain.  For this  purpose, we divide the  system in two
parties: one  consists of  just one qubit  and the other  contains the
remaining  $L-1$ qubits.   Then, following  Eq.~(\ref{eq:entropy}), we
compute $S_1$ from  the $2\times 2$ reduced density  matrix of the one
qubit subsystem.   In Fig.~\ref{fig:S1}  the behavior of  $S_1$ across
the transition  to chaos is shown  for the AA model  and for different
sizes of the system, from $L=6$ to $L=12$.  We find that the bipartite
entanglement $S_1$  shows the same behavior independently  of the size
of  the system.  The  state of  the system  changes from  separable to
maximally entangled as the transition  to chaos occurs.  In all cases,
the entropy saturates to its  maximum value $S_1=1$, up to corrections
of    order   $1/2^L$    \cite{random-states}.    As    discussed   in
Sec.~\ref{sec:measures}, these results show that there exists a global
entanglement of each single qubit with the rest of the system and that
this entanglement increases with the interaction. The maximum value of
bipartite entanglement is obtained when  quantum chaos has set in.

We have obtained similar  results for the bipartite  entanglement when
the two  blocks  in  which the  system  is  partitioned have different
lengths.  As  an  example  we  show  in  Fig.~\ref{fig:S_vs_L} the Von
Neumann entropy $S_n$ between a block of size  $n$ and the rest of the
system (of size $L-n$)  from $n=1$ (bottom)  to $n=5$ (top) for the AA
model and a chain of  size $L=12$. Similarly  to the $S_1$ case, $S_n$
increases when the transition  to chaos occurs  and saturates to  $S_n
\approx n$ for  large interaction strength.  Hence,  the state of  the
system becomes  maximally entangled  when  chaos sets in.  This  is  a
direct consequence of   the  existence of   multipartite entanglement.
Moreover,  in the inset of Fig.~\ref{fig:S_vs_L},  we have plotted the
saturation  value of $S_n$ for  $J/J_{\mathrm c}=15$  as a function of
the size of subsystems $n$.  This shows that in the chaotic regime the
bipartite entanglement  scales linearly with the  size of the smallest
of  the two  blocks in which  the global  system has been partitioned:
$S_n \propto n$.

\begin{figure}[!t]
\begin{center}
\includegraphics[scale=0.4]{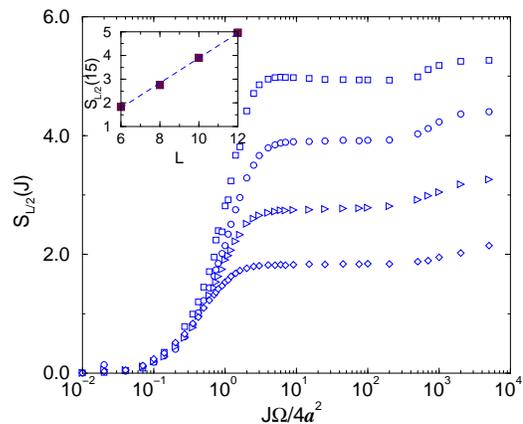}
\caption{\label{fig:S_half}
Von Neumann entropy  $S_{L/2}$ as a  function  of the strength of  the
interaction $J$, averaged over  all eigenfunctions in the central band
of the spectrum and over 10  random realizations.  The data correspond
to the  AA model  of Eq.~(\ref{eq:1D-V})  for a chain  of sizes  $L=6$
(diamonds), $L=8$ (triangles), $L=10$ (circles), $L=12$ (squares).  In
the  inset $S_{L/2}(J/J_{\mathrm c}=15)$ is plotted   as a function of
$L$.  The  dashed  line corresponds to a  linear  fit with a  slope of
$0.52 \pm 0.02$.}
\end{center}
\end{figure}

It is interesting  to study the Von Neumann  entropy as a  function of
the system size   $L$, when the two   blocks  in which the  system  is
partitioned have  a size $\propto L$. We   have computed the bipartite
entanglement $S_{L/2}$ corresponding to  the case in which the  system
is partitioned  in two halves.  A value  of $S_{L/2}>0$ for any $L$ is
indicative of   the  existence  of  multipartite  entanglement.    The
obtained behavior of $S_{L/2}$  as a function  of the strength of  the
interaction is   shown in Fig.~\ref{fig:S_half}, for  different system
sizes.  The behavior of  $S_{L/2}$ is similar  to that shown by $S_1$.
It takes very small values in the integrable regime and then increases
with  the interaction  up   to a value  for which   it saturates.  The
saturation value    is   $\approx   L/2$   (up to    corrections    of
$\mathcal{O}(1)$  \cite{random-states}).       In    the  inset     of
Fig.~\ref{fig:S_half},   we   plot    the   value  of $S_{L/2}$    for
$J/J_{\mathrm  c}=15$ ({\it i.e.}, in the  chaotic regime in which the
eigenstates in the central  band are effectively  mixed) as a function
of the size $L$ of the system.  It is interesting to note that the Von
Neumann  entropy does  feel the   mixing of  different spectral  bands
occurring   for  a  very  strong  interaction   ($J/J_{\mathrm c} \sim
1000$). The  inter-band mixing (compare  with Fig.~\ref{fig:1D-PN} for
$L=10$)   produces  a  increase  in the    Von   Neumann entropy which
nevertheless is  small compared  to  that observed  for the  onset  of
chaos.
This is   in   contrast with the    pairwise  measures  such  as   the
concurrence, for which we did not observed any change.

In  addition, for any  given value of $J$  when chaos has  set in, the
bipartite entanglement scales linearly  with  the size of the  system.
It is  interesting  to  comment this   result  from the viewpoint   of
computational complexity.  It was
shown in \cite{vidal} that large entanglement  of the quantum computer
hardware  is a  necessary   condition  for exponential speedup   (with
respect to classical computation) in  quantum computation operating on
pure  states.  To  be more   precise,   a necessary condition for   an
exponential  speedup is that   the  amount of  entanglement  increases
greater than logarithmically  with  the size  $L$  of the computation.
This  condition    is    fulfilled  in  the   chaotic    regime  where
$S_{L/2}\propto  (L/2)$.  We remark  that,  differently from  problems
like exact  cover \cite{Orus}, this  is not limited to  the transition
region but extends to the whole chaotic regime.
We also note that the relation between entanglement and computational
complexity in quantum algorithms simulating quantum chaos
has been investigated in Ref.~\cite{montangero}.

\subsection{Weak and hard chaos}
\label{sec:discussion}

In this  section, we discuss  the behavior of quantum  entanglement in
situations of  weak and of  hard chaos.  As  it was discussed,  the NN
model  of Eq.~(\ref{eq:1D-V-NN}),  while similar  in character  to the
model of  Eq.~(\ref{eq:1D-V}) with  $l_c>1$, shows a  quite unexpected
peculiarity:   The   chaos  border   does   not   coincide  with   the
delocalization border.  Thus, when  the strength of the interaction is
increased, the NN model experiences a transition from integrability to
a situation of weak chaos  in which the eigenfunctions are delocalized
while the level statistics is  yet of Poissonian nature.  This results
from the fact that the NN Hamiltonian can be approximately mapped into
a  model of $L$  free fermions  as discussed  in \cite{free-fermions}.
However, this non generic situation is removed if longer ranges of the
interaction are considered.  This  gives us the possibility to compare
the behavior of entanglement in situations of weak and hard chaos.

\begin{figure}[!t]
\begin{center}
\includegraphics[scale=0.39]{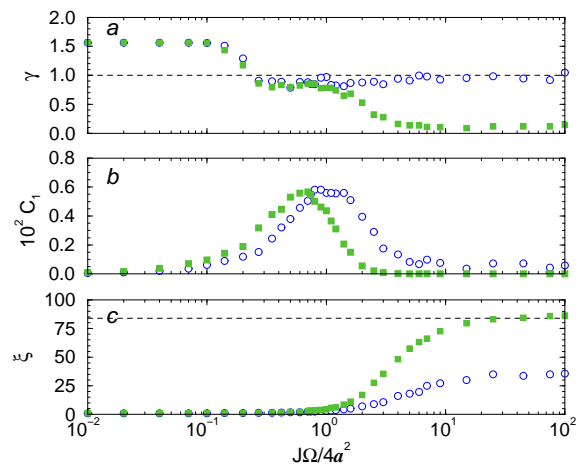}
\caption{\label{fig:chaos_vs_loc}  Comparison  between  the  NN  model
Eq.~(\ref{eq:1D-V-NN}) (circles) and  the model of Eq.~(\ref{eq:1D-V})
with $l_c=L/2$  (squares), for  a chain of  qubits  of length  $L=10$.
>From  top  to bottom, the  three panels   show, as  a  function of the
coupling  parameter $J$:  the    behavior of the   spectral statistics
parameter  $\gamma$ (top  panel);  the mean concurrence  $C_1$ between
nearest  neighbor pairs of  qubits  (middle panel); the  participation
number $\xi$ (bottom panel).  All   quantities were calculated in  the
central band  of  the    spectrum   and  averaged  over     30  random
realizations.}
\end{center}
\end{figure}

We have calculated the nearest  neighbor concurrence $C_1$ for the  NN
model  and for a long   range interaction model   with $l_c=L/2$.   In
Fig.~\ref{fig:chaos_vs_loc} we present  our results.  In panels  ($a$)
and   ($c$) the  $\gamma$   parameter and  the  PN   $\xi$  are shown,
respectively.   We  can clearly see the  peculiar  behavior  of the NN
model.   For  both  NN (open circles)   and  $l_c=L/2$ (solid squares)
models the  PN behaves  in a similar   way. For  the  NN model  the PN
signals a clear  transition from localized to delocalized eigenstates,
even thought  it  does not saturates  at  the  value corresponding  to
Gaussian  fluctuations.  In  particular,  the delocalization border of
both  models coincide.  Nevertheless,  the  $\gamma$ parameter shows a
very    different   behavior.   For   interaction   strengths  $0.2\le
J/J_{\mathrm c}
\le 1$ the level statistics  parameter $\gamma$ for both models  takes
the value $\gamma \approx 1$.  This corresponds to the integrable case
in which  the  nearest neighbor  spacing distribution  is  Poissonian.
However, for  the  NN model,  $\gamma$ remains Poissonian   for larger
values of the interaction up to $J/J_{\mathrm c}\approx 100$.  This is
the situation that  has been termed as weak  chaos.   In contrast, for
the   $l_c=L/2$  model the  level  statistics  changes from Poissonian
($\gamma=1$), to GOE ($\gamma=0$),  around $J/J_{\mathrm c}=1$.  Thus,
for   the  $l_c=L/2$ model   the  chaos   border  coincides  with  the
delocalization border as it is commonly found in many-particle systems
with two-body interaction.

In panel ($b$) the corresponding results for the concurrence $C_1$ are
shown.  We observe  again the difference in  $C_1$ due to the range of
the interaction,  as discussed  in the previous   subsection.  Despite
this difference, $C_1$ shows a similar behavior for both models: It is
small at both sides of the transition and increases in between, having
its maximum value   close   to $J/J_{\mathrm  c}=1$.   This  numerical
results suggest that the behavior of the pairwise entanglement is more
sensitive to  the mixing   of the eigenstates   than to  the  onset of
quantum chaos.

\section{Final Remarks}
\label{sec:conclusions}

We have studied the bipartite and pairwise entanglement in one and two
dimensional spin  lattice  models  that  experience a   transition  to
quantum chaos.

To  study the presence of  multipartite entanglement, we have analyzed
the behavior  of the averaged  Von Neumann  entropy  for subsystems of
different sizes.  In particular,  we have shown  that, for a partition
of the system into   two  equal-size subsystems, this  quantity  grows
linearly with the system size in the  chaotic regime.  This shows that
the  classical simulation method discussed   in \cite{vidal} cannot be
used  to efficiently simulate the  quantum chaos regime on a classical
computer.

For the case of pairwise entanglement,  we have studied the dependence
of the concurrence on the distance between the  partners, the range of
the interaction and the size  of the system.  Our results suggest that
for a      typical many-qubit state,      the entanglement  is  mainly
multipartite rather than pairwise. 

We have also discussed  the   different character that  the   pairwise
entanglement has at  the   integrable  and the  chaotic  side   of the
transition in terms  of a suitable distribution  of the eigenvalues of
the two-spin  reduced density matrix. The  use of  the moments of this
distribution  to  mark the   transition to  quantum  chaos remains and
interesting open question.

Finally,  we have discussed  the similarities  and differences between
the behavior of the concurrence at a  quantum phase transition and at
the    onset of  quantum chaos.  Our    results show that  the maximal
concurrence  is obtained close to  the delocalization border for which
mixing  of the    noninteracting eigenfunctions takes   place  and not
necessarily related to the onset of quantum chaos.

\vspace{1cm}

This work was supported in part by the EC contract IST-FET EDIQIP, the
NSA and ARDA  under ARO contract No.   DAAD19-02-1-0086,  and the PRIN
2002 ``Fault tolerance,  control and stability  in quantum information
processing''.  We acknowledge useful discussions  with B. Georgeot, F.
Izrailev, D. Shepelyansky,  T.H.  Seligman, and  V.  Sokolov.  C.M.-M.
acknowledges the  hospitality of ``Centro  Internacional de Ciencias''
where part of this work has been done.



\end{document}